\documentstyle[preprint,aps,12pt,floats]{revtex}

\newcommand{\sth}{\sin\beta}
\newcommand{\sthz}{\sin^{2}\beta}
\newcommand{\szth}{\sin 2\beta}

\newcommand{\cth}{\cos\beta}
\newcommand{\cthz}{\cos^{2}\beta}
\newcommand{\schi}{\sin\gamma}
\newcommand{\cchi}{\cos\gamma}

\newcommand{\szchi}{\sin 2\gamma}
\newcommand{\czchi}{\cos 2\gamma}

\newcommand{\gva}{\gamma_{VA}}
\everymath={\displaystyle}
\begin{document}
%%%%%%%%%%%%%%%%%%%%% TITLE PAGE %%%%%%%%%%%%%%%%%%%%%%%%%%%%%%%%%%%%%%%%%%%%%%
%
\preprint{
\font\fortssbx=cmssbx10 scaled \magstep2
\hbox to \hsize{
%\special{psfile=uwlogo.ps
%hscale=8000 vscale=8000
%hoffset=-12 voffset=-2}
%\hskip.5in \raise.1in\hbox{\fortssbx University of Wisconsin - Madison}
\hfill$\vtop{   \hbox{\bf MADPH-95-906\\}
                \hbox{\bf HUTP-95/A030\\}
                \hbox{\bf hep-ph/9508312\\}
                \hbox{August 1995}}$ }
}
\title{
Decay Rates, Structure Functions and
New Physics Effects in  Hadronic Tau Decays
}
\author{Markus Finkemeier}
\address{ Lyman Laboratory of Physics, Harvard University, Cambridge, MA 02148,
USA}
\author{Erwin Mirkes\thanks{Invited talk presented by E. Mirkes
at the Workshop on the Tau/Charm Factory, Argonne National Laboratory,
June 21-23, 1995}
}
\address{
Department of Physics, University of Wisconsin, Madison, WI 56706, USA}
\maketitle
\begin{abstract}
Hadronic decays rates  of the $\tau$ lepton into
multi meson final states are presented.
The structure of the hadronic matrix elements
for various decay modes is discussed.
The formalism of structure functions
allows for a detailed test of these matrix elements.
Various correlations are discussed which are sensitive to
possible CP violation and new physics effects in the
decay modes.
\end{abstract}
%
%%%%%%%%%%%%%%%%%%%%%%%%%%%%%%%%%%%%%%%%%%%%%%%%%%%%%%
\section{INTRODUCTION}
%%%%%%%%%%%%%%%%%%%%%%%%%%%%%%%%%%%%%%%%%%%%%%%%%%%%%%
The $\tau$ lepton is heavy enough to decay into a variety of hadronic
final states.
In particular, final states with kaons provide a powerful
probe of the strange sector of the weak charged current.
The Tau-Charm Factory operating at an $e^+e^-$ cms energy of around
4 GeV  and a luminosity of $L=10^{33}\mbox{cm}^{-2}\mbox{s}^{-1}$ with
good $\pi/K$ separation
\cite{kirkby} would allow for  high precision measurements of the
hadronic matrix elements in all decay modes.
Rare decay modes could be searched for at the level of about $10^{-7}$
in branching fraction. Of particular interest would also be the search
for possible  CP violation in the hadronic matrix elements.

In the present paper, we specify the general structure of
the matrix elements for $\tau$ decays into various multi meson final states.
We study  angular correlations in the exclusive decay modes
and show that the formalism of structure functions allows for a
detailed model independent test of the hadronic matrix elements.
Furthermore, the structure functions
allow for a systematic analysis of possible
CP violation effects in the matrix elements, which would have to come
from new non-Standard Model contributions.

It is shown  that CP violation
effects  are in principle
observable in a Tau-Charm Factory (without polarized beams)
for three meson decay modes
with a nonvanishing vector  {\it and}
an axial vector current.
CP violation effects originating from a charged Higgs could  be detected
only for decay modes with a nonvanishing vector current.

An observation of CP violation in two meson decays requires either
polarized beams \cite{tsaicp}
or kinematical information from the second tau decay \cite{nelson1}.

%%%%%%%%%%%%%%%%%%%%%%%%%%%%%%%%%%%%%%%%%%%%%%%%%%%%%%
\section{MATRIX ELEMENTS AND DECAY RATES }
%%%%%%%%%%%%%%%%%%%%%%%%%%%%%%%%%%%%%%%%%%%%%%%%%%%%%%
The  matrix element ${\cal{M}}$  for the hadronic $\tau$ decay
into $n$ mesons $h_1, \ldots h_n$
\begin{equation}
\tau(l,s)\rightarrow\nu(l^{\prime},s^{\prime})
+h_{1}(q_{1},m_{1})+\ldots h_{n}(q_{n},m_{n}) \>,
\label{process2h}
\end{equation}
can be expressed in terms of a leptonic ($M_\mu$) and a
hadronic  current  ($J^\mu$) as
\begin{equation}
{\cal{M}}=\frac{G}{\sqrt{2}}\,
\bigl(^{\cos\theta_{c}}_{\sin\theta_{c}}\bigr)
\,M_{\mu}J^{\mu} \>.
\label{mdef2h}
\end{equation}
In Eq.~(\ref{mdef2h}),
$G$ denotes the Fermi-coupling constant and   $\theta_c$ is the
Cabibbo angle. The leptonic current is given by
\begin{equation}
M_{\mu}=
\bar{u}(l^{\prime},s^{\prime})\gamma_{\mu}(g_V-g_A\gamma_{5})u(l,s) \>,
\label{leptoncurrent}
\end{equation}
with $g_V=g_A=1$ in the Standard Model.
The hadronic current $J^{\mu}$ can  in general be  expressed in terms of
a vector and an axial vector current
\begin{equation}
J^{\mu}(q_{1},\ldots,q_{n})=\langle h_{1}(q_{1})\ldots h_{n}(q_n)
|V^{\mu}(0)-A^{\mu}(0)|0\rangle \>.
\label{hadmat2h}
\end{equation}
The simplest decay mode into a pion or a kaon proceeds only through the
axial vector current whereas all decays into an even number of pions
are expected to proceed through the vector current. In fact, the decay
rates for $\tau\rightarrow 2n\pi, KK$ can be related through the
conserved vector current (CVC) hypothesis  to $e^+e^-\rightarrow$
hadrons in the isovector state \cite{tsai1}.
On the other hand, three body decay modes involving kaons
 allow for axial and vector current contributions at
the same time.
In the following, we specify  the hadronic matrix elements
for hadronic decays into multi meson final states as expected from the
Standard Model.

%%%%%%%%%%%%%%%%%%%%%%%%%%%%%%%%%%%%%%%%%%%%%%%%%%%%%%
\subsection{One Meson Decays }
%%%%%%%%%%%%%%%%%%%%%%%%%%%%%%%%%%%%%%%%%%%%%%%%%%%%%%

The decay rate for the simplest decay mode with one pion or kaon
is well
predicted by the the pion or kaon kaon decay constants $f_{\pi}$  and $f_K$
defined by the matrix element of the axial vector currents
\begin{eqnarray}
\langle \pi(q) |A^{\mu}(0)|0\rangle \> &=& i\sqrt{2}f_{\pi} q^{\mu}\\
\langle K(q) |A^{\mu}(0)|0\rangle \>   &=& i\sqrt{2}f_{K} q^{\mu}.
\end{eqnarray}
Both decay constants can be determined using the precisely measured
pion (kaon) decay widths $\Gamma(\pi (K)\rightarrow \mu \nu_\mu)$.
Radiative corrections $\delta R_{\tau/\pi}=(0.16\pm0.14)\%$
and $\delta R_{\tau/K}=(0.90\pm 0.22)\%$
to the ratios
$\Gamma(\tau\rightarrow\pi\nu)/\Gamma(\pi\rightarrow\mu\nu)$
and
$\Gamma(\tau\rightarrow K\nu)/\Gamma(K\rightarrow\mu\nu)$
have been calculated recently \cite{markus1}.
Using the recent world average $\tau_{\tau}=(291.6\pm1.6)\,\mbox{fs}$
for the tau lifetime  \cite{davier} one obtains the following
theoretical predictions for the branching ratios
\begin{eqnarray}
{\cal B}(\pi\nu_\tau) &=& (10.95\pm 0.06)\% \\
{\cal B}(K\nu_\tau)   &=& (0.723\pm 0.006)\%
\end{eqnarray}
These predictions agree within one standard deviation with the
world averages as quoted in \cite{Hel94}.

%%%%%%%%%%%%%%%%%%%%%%%%%%%%%%%%%%%%%%%%%%%%%%%%%%%%%%
\subsection{Two Meson Decays }
%%%%%%%%%%%%%%%%%%%%%%%%%%%%%%%%%%%%%%%%%%%%%%%%%%%%%%

The hadronic matrix element for the decay
$\tau\rightarrow h_1 h_2 \nu_{\tau}$
can be written as ($Q^{\mu}=(q_{1}+q_{2})^{\mu}$)
\begin{equation}
\langle h_1(q_1) h_2(q_2) |V^{\mu}(0)|0\rangle \> =
[ (q_1-q_2)_\nu\,T^{\mu\nu}\,  F^{h_1 h_2} + Q^\mu\,F^{h_1 h_2}_4]
\label{mat2h}
\end{equation}
$T^{\mu\nu}$ is the transverse projector, defined by
\begin{equation}
T_{\mu\nu}=  g_{\mu \nu} - \frac{Q_\mu Q_\nu}{Q^2}  \>.
\label{trans}
\end{equation}
The form factor $F_4$ describes the two mesons $h_1$ and $h_2$
in an $s$ wave.
As mentioned before, the form factor $F^{\pi\pi}$ in
$\tau^-\rightarrow \rho^-\nu\rightarrow \pi^-\pi^0\nu$
can be obtained (using the CVC theorem)
from the iso-vector part of the electromagnetic current
for $e^+e^-\rightarrow\pi^+\pi^-$  and
%the contribution to
the scalar form factor $F_4$ is expected to
vanish.
One has
\begin{equation}
F^{\pi^-\pi^0} = \sqrt{2} \,T_\rho^{(1)}\>,
\end{equation}
where
where $T_\rho^{(1)}$ is a normalized vector resonance form factor (two
particle Breit-Wigner propagator) for the $\rho$ resonance
including  the contribution from the radial excitations $\rho^\prime$
and $\rho^{\prime\prime}$.
In general, the  normalization of the form factors $F^{h_1h_2}$ is fixed by
chiral symmetry constraints, which determines the matrix elements in
the limit of soft meson momenta. The strong interaction effects beyond
the low energy limit are taken into account by vector resonance
factors with the requirement $T_{X}^{(1)}(Q^2=0)=1$ ($X = \rho$, $K^\star$).
The hadronic matrix elements for
the Cabibbo suppressed decay modes
$K^-\pi^0\nu_\tau,\, \overline{K^0}\pi^-\nu_\tau$
are dominated by the  $K^\star$ resonance $T_{K^\star}^{(1)}(Q^2)$ \cite{GR},
whereas the one for the Cabibbo allowed mode $K^0 K^- $
is dominated by the high energy tail of the $\rho$.
One has \cite{fm1}.

\begin{eqnarray}
F^{\overline{K^0} \pi^-}    &=&
       \frac{1}{\sqrt{2}}T_{K^\star}^{(1)}(Q^2) \>,        \\
F^{K^- \pi^0}     &=&  \hspace{5mm} T_{K^\star}^{(1)}(Q^2)   \>,        \\
F^{K^0 K^-}       &=&  \hspace{5mm} T_\rho^{(1)}(Q^2)        \>.
\end{eqnarray}
In the $\tau\rightarrow K\pi$ decay mode, $F_4$ gets a contribution
from the off-shellness $(m_{K*}^2 - Q^2)$ of the $K^\star$.
However, this scalar contribution is strongly suppressed compared to
the contribution of $F$.
As we will see in the last section, the form factor $F_4$ allows also
for a possible contribution from a charged Higgs exchange
and is therefore of special interest.

We use the following  form for the two particle
Breit-Wigner propagators with an energy dependent
width $\Gamma_X(s)$ throughout this paper:
\begin{equation}
\mbox{BW}_{X}[s]\equiv {M^2_X\over [M^2_X-s-i\sqrt s \Gamma_X(s)]}\>,
\end{equation}
where $X$ stands for the various resonances of the two meson
channels.
The following parametrization is used for the $\rho$ resonance:
\begin{eqnarray}
   T_\rho^{(1)}(s) & = & \frac{1}{1 + \beta_\rho}
  \Big[ \mbox{BW}_\rho(s) + \beta_\rho\, \mbox{BW}_{\rho'}(s)
  \Big] \>,
   \label{beta}
\end{eqnarray}
where
$
\beta_\rho = -0.145\>,
m_\rho = 0.773 \, \mbox{GeV}\>, \Gamma_\rho = 0.145 \,
m_{\rho'} = 1.370 \, \mbox{GeV}\>, \Gamma_{\rho'} = 0.510 \, \mbox{GeV}\>.
$
These are the values which have been determined from $e^+ e^- \to \pi^+
\pi^-$ in \cite{KueSa}.
The parameterization for $T_{K^\star}^{(1)}(Q^2)$ allows
for a contribution of the first excitation  ${K^\star}'(1410)$
in analogy to Eq.~(\ref{beta}):
\begin{eqnarray}
   T_{K^\star}^{(1)}(s) & = &\frac{1}{1 + \beta_{K^\star}}
  \Big[ \mbox{BW}_{K^\star}(s) + \beta_{K^\star}\, \mbox{BW}_{{K^\star}'}(s)
  \Big]\>,
  \label{betakst}
\end{eqnarray}
where
$
\beta_{K^\star} = -0.135\>,
m_{{K^\star}} = 0.892 \, \mbox{GeV}\>,
\Gamma_{{K^\star}} = \, 0.050 \mbox{GeV}\>,
m_{{K^\star}'} = 1.412\, \mbox{GeV}\>,
\Gamma_{{K^\star}'} = 0.227\, \mbox{GeV}\>.
$
The parameter   $\beta_{K^\star}$
was fixed in \cite{fm1}  by comparing the theoretical results
to the recent experimental branching ratio for
${\cal B}(K^\star\nu_\tau)= 1.36\pm 0.08$ \cite{Hel94}.
The value $\beta_{K^\star} = -0.135$
is remarkably close to the strength of  the
$\rho'$ contribution  to the $\rho$ Breit-Wigner,
supporting the use of approximate
$SU(3)$ flavour symmetry.

The branching ratios based on these parametrizations are
$
{\cal B}(\pi^- \pi^- \nu_\tau) = 23.5 \%\>,
{\cal B}(\overline{K^0} \pi^- \nu_\tau) = 0.45 \%\>,
{\cal B}(K^- \pi^0            \nu_\tau) = 0.9 \%\>,
$
For the decay into two kaons we obtain
$
   {\cal B}(K^0 K^- \nu_\tau) = 0.11 \%\>,
$
in good agreement with the recent world average
$   {\cal B}(K^0 K^- \nu_\tau) = 0.13 \pm 0.04 \% $ \cite{Hel94}.

%%%%%%%%%%%%%%%%%%%%%%%%%%%%%%%%%%%%%%%%%%%%%%%%%%%%%%
\subsection{Three Meson Decays }
%%%%%%%%%%%%%%%%%%%%%%%%%%%%%%%%%%%%%%%%%%%%%%%%%%%%%%

The hadronic matrix elements for three meson final states
have a much richer structure.
The decay modes involving  kaons allow
for axial and vector current contributions at the same time
\cite{braaten,Dec93}.
\begin{equation}
J^{\mu}(q_{1},q_{2},q_{3})=\langle h_{1}(q_{1})h_{2}(q_{2})h_{3}(q_{3})
|V^{\mu}(0)-A^{\mu}(0)|0\rangle \>.
\label{hadmat}
\end{equation}
The most general ansatz for the matrix element of the
quark current $J^{\mu}$  in Eq.~(\ref{hadmat})
is characterized by four form factors $F_i$ \cite{km1}, which
are in general functions of $Q^2$,
$s_1=(q_2+q_3)^2, s_2=(q_1+q_3)^2$ and $s_3=(q_1+q_2)^2$
\begin{eqnarray}
J^{\mu}(q_{1},q_{2},q_{3})
&=&     V_{1}^{\mu}\,F_{1}
    + V_{2}^{\mu}\,F_{2}
    +\,i\, V_{3}^{\mu}\,F_{3}
    + V_{4}^{\mu}\,F_{4} \>,
    \label{f1234}
\end{eqnarray}
with
\begin{equation}
\begin{array}{ll}
V_{1}^{\mu}&= (q_{1}-q_{3})_{\nu}\,T^{\mu\nu}  \>,\\
V_{2}^{\mu}&= (q_{2}-q_{3})_{\nu}\,T^{\mu\nu}  \>,\\
V_{3}^{\mu}&= \epsilon^{\mu\alpha\beta\gamma}q_{1\,\alpha}q_{2\,\beta}
                                             q_{3\,\gamma} \>,
\\
V_{4}^{\mu}&=q_{1}^{\mu}+q_{2}^{\mu}+q_{3}^{\mu}\,=Q^{\mu} \>.
    \end{array}
\label{videf}
\end{equation}
$T^{\mu\nu}$ denotes again the transverse projector as defined
 in Eq.~(\ref{trans}).
The form factors $F_{1}$ and $F_{2} (F_{3})$ originate from the axial vector
hadronic current (vector current) and correspond to a hadronic system
in a spin one state,
whereas $F_{4}$  is due to the spin zero part of the axial current matrix
element.
In the limit of vanishing quark masses, the weak axial-vector
current is conserved and this implies that the scalar form factor
$F_4$ vanishes. The massive pseudoscalars  give a
contribution to $F_4$, however, the effect is very small
\cite{dfm} and we will neglect this contribution in the subsequent
discussion of this section.
Note however that the form factor $F_4$ in the
$\tau\rightarrow(3\pi)\nu_\tau$ decay mode
could receive a sizable contribution due to the
$J^P=0^-$ resonance of the $\pi^\prime$ \cite{isgur,km1}.
Furthermore, the form factor $F_4$ allows also
for a possible contribution from a charged Higgs exchange.
We will consider this in more detail in the last section.

The form factors $F_1$  and $F_2$ can be predicted by chiral lagrangians,
supplemented by informations about resonance parameters.
Parametrizations for the $3\pi$ final states
based on this model can be found in \cite{KueSa,km1,km2}.
In this case the vector form factor is absent due to the
$G$ parity of the pions. On the other hand, the decay mode
$\tau^-\rightarrow \eta\pi^-\pi^0\nu_{\tau}$ has a vanishing
contribution from the axial vector current \cite{kramer,pich3,Dec93}.
The vector form factor is related
to the Wess-Zumino anomaly \cite{WZ,kramer}
whereas the axial-vector form factors are again
predicted by chiral Lagrangians
as mentioned before.
A general parameterization of the form factors for various three meson
decays modes with pions and kaons
was proposed in \cite{Dec93}.
The parameterization  has been extensively reanalyzed
in \cite{fm1} which lead to sizable differences in the
predictions of the decay rates compared to \cite{Dec93}.
Furthermore, a parameterization for the final states
with two neutral kaons
$\tau^-\rightarrow K_S\pi^-K_S\nu_\tau, \,
 \tau^-\rightarrow K_L\pi^-K_L\nu_\tau, $ and
$\tau^-\rightarrow K_S\pi^-K_L\nu_\tau$ was derived in \cite{fm1}.
The results for the form factors $F_i$ in Eq.~(\ref{f1234}) for the
decay modes $\tau\rightarrow abc\nu_\tau$ are
summarized by
\begin{eqnarray}
F^{(abc)}_{1}(Q^2,s_2,s_3)&=&{2\sqrt 2 A^{(abc)}\over 3f_\pi}
                              G_{1}^{(abc)}(Q^2,s_2,s_3) \>,
                              \label{f1}\\
F^{(abc)}_{2}(Q^2,s_1,s_3)&=&{2\sqrt 2 A^{(abc)}\over 3f_\pi}
                              G_{2}^{(abc)}(Q^2,s_1,s_3) \>,
                              \label{f2}\\
F^{(abc)}_{3}(Q^2,s_1,s_2,s_3) &=& {A^{(abc)}\over 2\sqrt 2\pi^2f^3_\pi}
                              G_3^{(abc)}(Q^2,s_1,s_2,s_3)\>.
                              \label{f3}
\end{eqnarray}
The Breit-Wigner functions $G_{1,2}$ ($G_3$) and the normalizations
$A^{(abc)}$ are listed in
Tab.~\ref{tab1} (\ref{tab2})
for the various decay modes.
Note that by convenient ordering of the mesons, the two body resonances
in $F_1$ ($F_2$) occur only in the variables $s_2,s_3$ ($s_1,s_3)$.

%%%%%%%%%%%%%%%%%%%%%%%%% TABLES %%%%%%%%%%%%%%%%%%%%%%%%%%%%%%%%%%%%%%%%%%%%%%
%
%
%
%%%%%%%%%% Table 1 %%%%%%%%
\begin{table}
\caption{Parameterization of the form factors
$F_1$ and $F_2$ in Eqs.~(\protect\ref{f1},\protect\ref{f2})
for the matrix elements of the weak axial-vector
current for the various channels.}
\label{tab1} \nopagebreak
$$
\begin{array}{c@{\quad}c@{\quad}c@{\quad}c}
\hline \hline
\begin{array}{c}
\mbox{channel} \\\mbox{(abc)}
\end{array} &
A^{(abc)} & G_1^{(abc)}(Q^2,s_2,s_3) &
G_2^{(abc)}(Q^2,s_1,s_3)
\\
\hline
\pi^- \pi^- \pi^+ &
\cos \theta_c &
\mbox{BW}_{A_1}(Q^2) T_\rho^{(1)}(s_2) &
\mbox{BW}_{A_1}(Q^2) T_\rho^{(1)}(s_1)
\\[1mm]
\pi^0 \pi^0 \pi^0 &
\cos \theta_c &
\mbox{BW}_{A_1}(Q^2) T_\rho^{(1)}(s_2) &
\mbox{BW}_{A_1}(Q^2) T_\rho^{(1)}(s_1)
\\[1mm]
K^- \pi^- K^+ &
\frac{- \cos \theta_c}{2} &
\mbox{BW}_{A_1}(Q^2) T_\rho^{(1)}(s_2) &
\mbox{BW}_{A_1}(Q^2) T_{K^\star}^{(1)} (s_1)
\\[3mm]
K^0 \pi^- \overline{K^0} &
\frac{- \cos \theta_c}{2} &
\mbox{BW}_{A_1}(Q^2) T_\rho^{(1)}(s_2) &
\mbox{BW}_{A_1}(Q^2) T_{K^\star}^{(1)} (s_1)
\\[3mm]
K_S \pi^- K_S &
\frac{- \cos \theta_c}{4} &
\mbox{BW}_{A_1}(Q^2) T_{K^\star}^{(1)}(s_3) &
%@@@
\begin{array}{c}
-\mbox{BW}_{A_1}(Q^2)
\times \\ {}
[ T_{K^\star}^{(1)} (s_1)
+ T_{K^\star}^{(1)}(s_3) ]
\end{array}
\\[4mm]
K_S \pi^- K_L &
\frac{- \cos \theta_c}{4} &
\begin{array}{c}
\mbox{BW}_{A_1}(Q^2)
\times \\{}
   [ 2 T_\rho^{(1)}(s_2) + T_{K^\star}^{(1)}(s_3)]
\end{array}
 &
\begin{array}{c}
\mbox{BW}_{A_1}(Q^2)\times \\{}
[ T_{K^\star}^{(1)} (s_1)
- T_{K^\star}^{(1)}(s_3) ]
\end{array}
\\[4mm]
K^- \pi^0 K^0 &
\frac{3 \cos \theta_c}{2 \sqrt{2}} &
\begin{array}{c}
\mbox{BW}_{A_1} (Q^2) \times \\{}
\left[ \frac{2}{3} T_\rho^{(1)}(s_2)
+ \frac{1}{3} T_{K^\star}^{(1)}(s_3) \right]
\end{array}
 &
\begin{array}{c}
\frac{1}{3}\mbox{BW}_{A_1} (Q^2) \times \\{}
\left[  T_{K^\star}^{(1)}(s_1)
-  T_{K^\star}^{(1)}(s_3) \right]
\end{array}
\\[4mm]  \hline
\pi^0 \pi^0 K^- &
\frac{\sin \theta_c}{4} &
T_{K_1}^{(a)} (Q^2) T_{K^\star}^{(1)}(s_2) &
T_{K_1}^{(a)} (Q^2) T_{K^\star}^{(1)}(s_1)
\\[3mm]
K^- \pi^- \pi^+ &
\frac{- \sin \theta_c}{2} &
T_{K_1}^{(a)} (Q^2) T_{K^\star}^{(1)}(s_2) &
T_{K_1}^{(b)} (Q^2) T_{\rho}^{(1)}(s_1)
\\[3mm]
\pi^- \overline{K^0} \pi^ 0 &
\frac{3 \sin \theta_c}{2 \sqrt{2}} &
\begin{array}{l}
\,\,\,\,\frac{2}{3} T_{K_1}^{(b)} (Q^2) T_\rho^{(1)} (s_2) \\[1ex]
+ \frac{1}{3} T_{K_1}^{(a)} (Q^2) T_{K^\star}^{(1)}(s_3)
\end{array} &
\begin{array}{c}
\frac{1}{3} T_{K_1}^{(a)} (Q^2) \times \\
\left[ T_{K^\star}^{(1)}(s_1)
 - T_{K^\star}^{(1)}(s_3) \right]
\end{array}
\\[8mm]  \hline \hline
\end{array}
$$
\end{table}

%%%%%%%%%% Table 2 %%%%%%%%

\begin{table}
\caption{Parameterization of the form factor
$F_3$ in Eq.~(\protect\ref{f3})
for the matrix elements of the weak vector current
for the various channels.}
\label{tab2}
$$
\begin{array}{c@{\quad}c@{\quad}c}
\hline \hline
\mbox{channel (abc)} & A^{(abc)} &
G_3^{(abc)}(Q^2,s_1,s_2,s_3)
\\
\hline
K^- \pi^- K^+ &
- \cos \theta_c &
T_\rho^{(2)}(Q^2) (\sqrt{2} - 1) \left[ \sqrt{2} T_\omega(s_2)
+ T_{K^\star}^{(1)}(s_1) \right]
\\ [1mm]
K^0 \pi^- \overline{K^0} &
\cos \theta_c &
T_\rho^{(2)}(Q^2) (\sqrt{2} - 1) \left[ \sqrt{2} T_\omega(s_2)
+ T_{K^\star}^{(1)}(s_1) \right]
\\[1mm]
K_S \pi^- K_S &
\frac{- \cos \theta_c}{2} &
T_\rho^{(2)}(Q^2) (\sqrt{2} - 1) \left[  T_{K^\star}^{(1)}(s_1)
- T_{K^\star}^{(1)}(s_3) \right]
\\[1mm]
K_S \pi^- K_L &
\frac{ \cos \theta_c}{2} &
T_\rho^{(2)}(Q^2) (\sqrt{2} - 1) \left[ 2 \sqrt{2} T_\omega(s_2)
+ T_{K^\star}^{(1)}(s_1)
+ T_{K^\star}^{(1)}(s_3) \right]
\\[1mm]
K^- \pi^0 K^0 &
\frac{- \cos \theta_c}{\sqrt{2}} &
T_\rho^{(2)}(Q^2) (\sqrt{2} - 1) \left[  T_{K^\star}^{(1)}(s_3)
- T_{K^\star}^{(1)}(s_1) \right]
\\[1mm]
\eta \pi^- \pi^0 &
\frac{ \cos \theta_c}{\sqrt{3}} &
T_\rho^{(2)}(Q^2) T_\rho^{(1)}(s_1)
\\[1mm]  \hline
\pi^0 \pi^0 K^- &
\sin \theta_c &
\frac{1}{4} T_{K^\star}^{(2)}(Q^2) \left[ T_{K^\star}^{(1)}(s_1)
- T_{K^\star}^{(1)}(s_2) \right]
\\[1mm]
K^- \pi^- \pi^+ &
 \sin \theta_c &
\frac{1}{2} T_{K^\star}^{(2)}(Q^2) \left[ T_\rho^{(1)}(s_1)
+  T_{K^\star}^{(1)}(s_2) \right]
\\[1mm]
\pi^- \overline{K^0} \pi^ 0 &
 \sqrt{2} \sin \theta_c &
\frac{1}{4} T_{K^\star}^{(2)}(Q^2) \left[2 T_\rho^{(1)}(s_2)
+ T_{K^\star}^{(1)}(s_1)
+ T_{K^\star}^{(1)}(s_3) \right]
 \\ \hline \hline
\end{array}
$$
\end{table}

%%%%%%%%%% Table 3 %%%%%%%%

\begin{table}
\caption{Predictions for the normalized decay widths
$\Gamma(abc)/\Gamma_e$ and the branching
ratios ${\cal B}(abc)$ for the various channels.
The contribution from the vector current is listed in column 3 and
available experimental data are listed in column 5.
The later are taken from \protect\cite{Hel94,Smi94,RPP94}.
}
\label{tab3}
$$
\begin{array}{c@{\quad}c@{\quad}c@{\quad}c@{\quad}c}
\hline \hline
\begin{array}{c}
\mbox{channel}
\\
(abc)
\end{array}
& \! \! \! \!
\left( \frac{\Gamma{(abc)}}{\Gamma_e} \right)^{(pred.)} & \! \! \! \!
\left( \frac{\Gamma{(abc)}}{\Gamma_e} \right)^{(pred.)}_{V} & \! \! \! \!
{\cal B}{(abc)}^{(pred.)}  & \! \! \! \!
{\cal B}{(abc)}^{(expt.)}
 \\
\hline
\\[-4mm]
\pi^- \pi^- \pi^+        & 0.48    & 0.  & 8.6  \%   & ( 8.64 \pm 0.24)  \% \\
\pi^0 \pi^0 \pi^-        & 0.48    & 0.  & 8.6  \%  &  (9.09 \pm 0.14)  \%
\\K^- \pi^- K^+          & 0.011   & 0.0045  & 0.20 \%  & ( 0.20 \pm 0.07)  \%
\\
K^0 \pi^- \overline{K^0} & 0.011   & 0.0045  & 0.20 \%  &                 \\
K_S \pi^- K_S            & 0.0027  & 0.0008  & 0.048 \% &  (0.021\pm 0.006) \%
\\
K_S \pi^- K_L            & 0.0058  & 0.0029  & 0.10 \%  &                 \\
K^- \pi^0 K^0            & 0.0090  & 0.0032  & 0.16 \%  &  (0.12 \pm 0.04) \%
\\
\eta \pi^- \pi^0         & 0.0108  & 0.0108  & 0.19 \%  &  (0.170 \pm 0.028) \%
 \\[1mm]
\hline \\[-4mm]
\pi^0 \pi^0 K^-          & 0.0080  & 0.0007  & 0.14 \%  &  (0.09 \pm 0.03) \%
\\
K^-   \pi^-\pi^+         & 0.043   & 0.0043  & 0.77 \%  &  (0.40 \pm 0.09) \%
\\
\pi^-\overline{K^0}\pi^0 & 0.054   & 0.0058  & 0.96 \%  &  (0.41 \pm 0.07) \%
\\[1mm]
\hline \hline
\end{array}
$$
\end{table}
Let us briefly  discuss the three particle resonances in Tab. I and
II (for details see \cite{fm1}).
We use the $A_1$ resonance  in the non-strange case  with energy dependent
width
%
%
%\begin{equation}
$
  \mbox{BW}_{A_1}(s) =
  \frac{m_{A_1}^2} {m_{A_1}^2 - s - i m_{A_1} \Gamma_{A_1} g(s) /
  g(m_{A_1})}\>,
$
%\end{equation}
%
%
with
$
m_{A_1}       = 1.251 \,\, \mbox{GeV}\>,
\Gamma_{A_1 } = \, 0.475\,\, \mbox{GeV}\>.
$
The  function $g(s)$ has been calculated in
\cite{KueSa}.
The three particle resonances with strangeness
are
\begin{eqnarray}
   T_{K_1}^{(a)}(s) & = & \frac{1}{1 + \xi}
   \Big[ \mbox{BW}_{K_1(1400)}(s) + \xi \mbox{BW}_{K_1(1270)} (s)
   \Big] \>,
\nonumber \\[2mm]
   T_{K_1}^{(b)}(s) & = & \mbox{BW}_{K_1(1270)}(s) \>.
\end{eqnarray}
with $\xi = 0.33$  \cite{fm1}.
The three body  vector resonances $T_\rho^{(2)}$
and $T_{K^\star}^{(2)}$
include the higher radial excitations
 $\rho'$ and $\rho''$
and  ${K^\star}'$ and ${K^\star}''$ \mbox{}
\begin{eqnarray}
   T_\rho^{(2)} & = & \frac{1}{1 + \lambda + \mu}
  \Big[ \mbox{BW}_\rho(s) + \lambda \mbox{BW}_{\rho'}(s)
  + \mu \mbox{BW}_{\rho''}(s)
  \Big] \>,
\nonumber \\[2mm]
   T_{K^\star}^{(2)} & = &\frac{1}{1 + \lambda + \mu}
  \Big[ \mbox{BW}_{K^\star}(s) + \lambda \mbox{BW}_{{K^\star}'}(s)
  + \mu \mbox{BW}_{{K^\star}''}(s)
  \Big]\>,
\end{eqnarray}
with  $\lambda =  -0.25,  \mu =  - 0.038 $.
The $\omega$ resonance
$
T_\omega(s) = \frac{1}{1 + \epsilon} [ \mbox{BW}_\omega(s) +
\epsilon \mbox{BW}_\Phi(s) ]
$
in the vector form factor $F_3$ in Tab. II
allows for a contribution of the $\phi$ with a relative strength
$\epsilon = 0.05\>$ \cite{fm1}.

Numerical results for the
hadronic decay widths $\Gamma{(abc)}$ normalized to the leptonic
width $\Gamma_e$
and for the branching ratios
in Tab.~\ref{tab3} based on this  model for the form factors.
The predictions for the branching ratios use
$\Gamma_e/ \Gamma_{\mbox{tot}}=17.8 \%$, as calculated from the experimental
values for the tau mass
$m_\tau = 1.7771 \, \mbox{GeV}$ and lifetime
$\tau_{\tau} = 291.6 \, \mbox{fs}$ \protect\cite{davier}.

%%%%%%%%%%%%%%%%%%%%%%%%%%%%%%%%%%%%%%%%%%%%%%%%%%%%%%
\subsection{Four Pion Decays }
%%%%%%%%%%%%%%%%%%%%%%%%%%%%%%%%%%%%%%%%%%%%%%%%%%%%%%
In order to predict the two tau decays into four pions,
$\tau^-\rightarrow \nu_{\tau} \pi^-\pi^-\pi^+\pi^0$ and
$\tau^-\rightarrow \nu_{\tau} \pi^0\pi^0\pi^0\pi^-$,
there are two possible approaches.

The first approach is based on the fact that
these tau decays are again related through CVC to
corresponding $e^+ e^-$ annihilation channels, namely to
$e^+ e^- \to 2\pi^+ 2\pi^-$ and $e^+ e^- \to \pi^+ \pi^- 2 \pi^0$.
And so by using the measured $e^+ e^-$ cross sections as input, the tau
decays can be predicted \cite{tsai1,GR,eidelman}, and the results are
in good agreement with the $\tau$ data \cite{Hel94,argus1,cleo1,bourdon}.
This approach, however, allows only to predict the integrated decay rates and
the four pion invariant mass distributions. In order to predict the various
two and three pion differential distributions, or in order to understand
angular distributions, a dynamical model is need.

Such a dynamical model has be constructed in \cite{markus2}
which uses the other possible approach. One follows along the lines which
have been used above to obtain the hadronic current in the
three meson modes.
Again one starts from the structure of the hadronic current in the chiral limit
and then implements low lying resonances in the various channels ($\rho$,
$\rho'$, $\rho''$, $A_1$ and $\omega$ mesons).
There are a few free parameters, which are fixed using the experimental
$e^+ e^- \to
2 \pi^+ 2 \pi^-$ cross sections and the measured decay rate of the $\tau\to
\omega \pi \nu_\tau$ sub-mode.
After parameter fixing, predictions for $e^+ e^- \to \pi^+ \pi^- 2 \pi^0$ and
for the four pion decay modes of the $\tau$ are obtained, including
detailed two, three and four pion differential mass distributions.
The various predictions agree well with the available experimental data.

The $\omega\pi$ contribution to the $4\pi$ final state
 is expected to proceed via
a vector current. However, a violation of $G$-parity would allow the
$\omega\pi$ system to be in an axial vector state, which could be
revealed by an analysis of the angular distribution in the $\omega\pi$
mode as introduced in \cite{dm1}

%%%%%%%%%% Table 4 %%%%%%%%

%%%%%%%%%%%%%%%%%%%%%%%%%%%%%%%%%%%%%%%%%%%%%%%%%%%%%%
\section{ANGULAR DISTRIBUTIONS AND STRUCTURE FUNCTIONS IN TWO AND THREE
MESON DECAY MODES}
%%%%%%%%%%%%%%%%%%%%%%%%%%%%%%%%%%%%%%%%%%%%%%%%%%%%%%
In this section, we study  angular distribution
of the hadronic system of two and three meson final states
which are accessible in a future $\tau$-charm factory.
We will assume
that the direction of the $\tau$ in the hadronic
rest frame is known and
that no spin informations of the decaying $\tau$ can be  used in the
analysis.
Of particular interest in the three meson case are the distributions
of the normal to the Dalitz plane and the distributions around this normal.
It is shown that the most general distribution in the three meson
case can be characterized by 16 structure functions most of which can
be determined under the conditions mentioned above.
The study of angular correlations
of the hadronic system  allows for much more detailed studies
of the hadronic charged current
than it is possible by rate measurements alone.
Special emphasis is put on $T$-odd triple momentum correlations,
which allow for the observation of $CP-$violating
contributions beyond the Standard Model.

%%%%%%%%%%%%%%%%%%%%%%%%%%%%%%%%%%%%%%%%%%%%%%%%%%%%%%
\subsection{Two Body Decays }
%%%%%%%%%%%%%%%%%%%%%%%%%%%%%%%%%%%%%%%%%%%%%%%%%%%%%%

Of particular interest in the two body decays is  the distribution
of the direction
of  $h_{1}\,$ ($\hat{q}_{1}=\vec{q}_{1}/|\vec{q}_{1}|$)
and the direction of the $\tau$ (denoted by $\vec{n}_{\tau}$) viewed
from the hadronic rest frame
%
%
%\begin{equation}
$
\cos\beta = \vec{n}_{\tau}\cdot \hat{q}_{1} \>.
$
%\end{equation}
%
%
After integration over the unobserved
neutrino direction,
the differential decay rate for a two meson final state
is given   by
\begin{eqnarray}
d\Gamma(\tau\rightarrow 2h)&=&
 \left\{
    \bar{L}_{B}{W}_{B}
  + \bar{L}_{SA}{W}_{SA}
  + \bar{L}_{SF}{W}_{SF}
  + \bar{L}_{SG}{W}_{SG}
    \right\}\times \label{gamma}\\[1mm]
&&
\hspace{-1.5cm}
\frac{G^{2}}{4m_{\tau}}
(g_{V}^{2}+g_{A}^{2})
\bigl(^{\cos^2\theta_{c}}_{\sin^2\theta_{c}}\bigr)
%\cos^{2}\theta_{c}\,
\frac{1}{(4\pi)^{3}}
\frac{(m_{\tau}^{2}-Q^{2})^{2}}{m_{\tau}^{2}}\,
\,|\vec{q}_{1}|\,\,
     \frac{dQ^{2}}{\sqrt{Q^{2}}}\,
     \frac{d\cos\beta}{2}\nonumber
\end{eqnarray}
with
%
%
%\begin{equation}
$
\vec{q}_{1}^{z}=\frac{1}{2\sqrt{Q^{2}}}\left(
[Q^{2}-m_1^{2}-m_2^{2}]^2-4m_{1}^{2}m_{2}^{2}\right)^{1/2}\>.
$
%\end{equation}
%
%
The hadronic structure functions $W_X$ can be expressed in terms of
the form factors $F$ and $F_4$  as defined in Eqs.~(\ref{mat2h}) as follows:
\begin{eqnarray}
W_B    &=& 4 (\vec{q}_1)^2\,|F|^2\,           \\
W_{SA} &=& Q^2\,  |F_4|^2\,                   \\
W_{SF} &=& 4\sqrt{Q^2}|\vec{q}_1|
          \, \mbox{Re}\left[FF_4^*\right] \\
W_{SG} &=&-4\sqrt{Q^2}|\vec{q}_1|\,
\mbox{Im}\left[FF_4^*\right]   \label{wsg}
\end{eqnarray}
The leptonic coefficients are
  \begin{equation}
  \begin{array}{lcrl}
\bar{L}_{B} &=&     &K_1 \sthz +  K_2         \\
\bar{L}_{SA}&=&     &K_2\\
\bar{L}_{SF}&=&  -   &K_2 \cth\\
\bar{L}_{SG}&=&     & 0
  \end{array}
  \end{equation}
with
\begin{equation}
K_1 =  1- ({m_\tau^{2}}/{Q^{2}});\,\,\,\,\,
K_2 =  ({m_\tau^{2}}/{Q^{2}});
\label{kdef}
\end{equation}
Note that the coefficient $\bar{L}_{SG}$ vanishes, if only the $\beta$
dependence of the decay is analyzed. In the case of a polarized $\tau$
(as it is the situation at LEP) one can use the
direction of the $\tau$ spin-vector $\vec{s}$ in the lab to define a further
angle $\alpha$
by $\cos\alpha=\frac{(\vec{n}_{\tau}\times\vec{s})\cdot
   (\vec{n}_{\tau}\times\hat{q}_{1})}{
   |\vec{n}_{\tau}\times\vec{s}|\,\,|\vec{n}_{\tau}\times\hat{q}_1|}
\>$ (see also Fig. 5 in \cite{km1}).
Taking into account the distribution with respect to this
angle would allow to measure also the structure function
$W_{SG}$.
Note that the structure function $W_{SG}$ is
proportional to the imaginary part of
the form factors $(FF_4^*)$ and requires nontrivial phases of the
amplitudes resulting from final state interactions. These strong
interaction phases are essential for the observation of possible CP
violation effects in the hadronic decay amplitudes.
However, in our case
the angle $\alpha$ is not observable and has
to be averaged out. Hence, the $T$-odd correlation
$\bar{L}_{SG}W_{SG}$ vanishes and no test of CP violation is possible.
However, a nonvanishing contribution to the
distributions
$\bar{L}_{SA}{W}_{SA}$ or $\bar{L}_{SF}{W}_{SF}$
would be a clear signal of a scalar contribution (parametrized by
$F_4$)  to the two meson decay modes.

%%%%%%%%%%%%%%%%%%%%%%%%%%%%%%%%%%%%%%%%%%%%%%%%%%%%%%
\subsection{Three Body Decays }
%%%%%%%%%%%%%%%%%%%%%%%%%%%%%%%%%%%%%%%%%%%%%%%%%%%%%%
Like in the two body case,
the three meson decay modes are  most easily analyzed in the
hadronic rest frame $\vec{q_1}+\vec{q_2}+\vec{q_3}=0$.
The orientation of the hadronic
system is in general  characterized by
three Euler angles ($\alpha,\beta$ and $\gamma$) as introduced in
\cite{km1,km2}.
Performing the analysis of $\tau \to \nu_\tau $+ 3 mesons in the
hadronic rest frame  has the advantage that the product of the
hadronic
and the leptonic tensors reduce to a sum \cite{km1}
$ L^{\mu\nu}H_{\mu\nu}=\sum_{X} \bar{L}_XW_X$.
In this system the hadronic tensor $H_{\mu\nu}$ is decomposed into 16
hadronic structure
functions $W_{X}$
corresponding to 16 density matrix elements for a hadronic system in a spin one
[contributions proportional to
$V_1^{\mu}F_1, V_2^{\mu}F_2, V_3^{\mu}F_3$ in Eq.(\ref{videf})]
and spin zero state $[V_4^{\mu}F_4]$
(nine of them originate from a pure spin one and the
remaining are pure spin zero  or interference terms).
The 16 structure functions contain the
dynamics of the three meson decay and
  depend only   on the hadronic invariants $Q^2$ and
the Dalitz plot variables $s_{i}$.
The leptonic factors
 $\bar{L}_X$ factorize the dependence on the Euler angles and
also depend on the chirality parameter
$\gamma_{VA}= \frac{2g_{V}g_{A}}{g_{V}^{2}+g_{A}^{2}}$.
In our case, one can measure two Euler angles $\beta$ and $\gamma$
defined by
$
\cos\beta=\vec{n}_{\tau}\cdot\vec{n}_{\perp} \>,
\cos\gamma=-\frac{\vec{n}_{\tau}\cdot\hat{q}_{3}}{
             |\vec{n}_{\tau}\times\vec{n}_{\perp}|}\>,
\sin\gamma=\frac{(\vec{n}_{\tau}\times\vec{n}_{\perp})\cdot\hat{q}_{3}}{
             |\vec{n}_{\tau}\times\vec{n}_{\perp}|}
               \>.
$
The vector $\vec{n}_\tau$ denotes the $\tau$ direction in the hadronic
rest frame.
The $(x,y)$ plane is aligned with the hadron momenta, {\it i.e.}
$\vec{n}_{\perp}=(\vec{q}_{1}\times\vec{q}_{2})/
                |\vec{q}_{1}\times\vec{q}_{2}|$ \,
(the normal to the hadronic plane )
pointing along $Oz$.
The $Ox$ axis  is defined by the
                direction of $\hat{q}_{3}=\vec{q}_{3}/|\vec{q}_{3}|$.
In the three pion case $\pi^{-}\pi^{-}\pi^{+}$ we choose
$\vec{q}_{3}=\vec{q}_{\pi^{+}}$ and $|\vec{q}_{2}|>|\vec{q}_{1}|$.

The differential decay rate with respect to these two angels
is then given by
\begin{eqnarray}
d\Gamma(\tau\rightarrow 3h)&=&
           \frac{G^{2}}{2m_\tau}
\bigl(^{\cos^2\theta_{c}}_{\sin^2\theta_{c}}\bigr)
 \left\{\sum_{X}\bar{L}_{X}W_{X}\right\}
\times \label{diffrat}\\[2mm]
&&\frac{1}{(2\pi)^{5}}\frac{1}{64}
\frac{(m_\tau^{2}-Q^{2})^{2}}{m_\tau^{2}}\,
     \frac{dQ^{2}}{Q^{2}}\,ds_{1}\,ds_{2}
     \,\frac{d\gamma}{2\pi}\,
      \frac{d\cos\beta}{2}\>.
      \nonumber
\end{eqnarray}
The leptonic coefficients $\bar{L}_X$ will be discussed below.
The dependence of the structure functions on the form factors $F_i$
reads \cite{km1}:
  \begin{eqnarray}  \hspace{3mm}
W_{A}  &=&   \hspace{3mm}(x_{1}^{2}+x_{3}^{2})\,|F_{1}|^{2}
           +(x_{2}^{2}+x_{3}^{2})\,|F_{2}|^{2}
           +2(x_{1}x_{2}-x_{3}^{2})\,\mbox{Re}\left(F_{1}F^{\ast}_{2}\right)
                                   \nonumber \\[1mm]
W_{B}  &=& \hspace{3mm} x_{4}^{2}|F_{3}|^{2}
                                   \nonumber \\[1mm]
W_{C}  &=&  \hspace{3mm} (x_{1}^{2}-x_{3}^{2})\,|F_{1}|^{2}
           +(x_{2}^{2}-x_{3}^{2})\,|F_{2}|^{2}
           +2(x_{1}x_{2}+x_{3}^{2})\,\mbox{Re}\left(F_{1}F^{\ast}_{2}\right)
                                   \nonumber \\[1mm]
W_{D}  &=&  \hspace{3mm}2\left[ x_{1}x_{3}\,|F_{1}|^{2}
           -x_{2}x_{3}\,|F_{2}|^{2}
           +x_{3}(x_{2}-x_{1})\,\mbox{Re}\left(F_{1}F^{\ast}_{2}\right)\right]
                                   \nonumber \\[1mm]
W_{E}  &=& -2x_{3}(x_{1}+x_{2})\,\mbox{Im}\left(F_{1}
                    F^{\ast}_{2} \right)\nonumber \\[1mm]
W_{F}  &=&  \hspace{3mm}
          2x_{4}\left[x_{1}\,\mbox{Im}\left(F_{1}F^{\ast}_{3}\right)
                     + x_{2}\,\mbox{Im}\left(F_{2}F^{\ast}_{3}\right)\right]
                                   \nonumber \\[1mm]
W_{G}  &=&- 2x_{4}\left[x_{1}\,\mbox{Re}\left(F_{1}F^{\ast}_{3}\right)
                     + x_{2}\,\mbox{Re}\left(F_{2}F^{\ast}_{3}\right)\right]]
                                   \nonumber \\[1mm]
W_{H}  &=& \hspace{3mm}
      2x_{3}x_{4}\left[\,\mbox{Im}\left(F_{1}F^{\ast}_{3}\right)
                     -\,\mbox{Im}\left(F_{2}F^{\ast}_{3}\right)\right]
                                   \label{wi} \\[1mm]
W_{I}  &=&- 2x_{3}x_{4}\left[\,\mbox{Re}\left(F_{1}F^{\ast}_{3}\right)
                     -\,\mbox{Re}\left(F_{2}F^{\ast}_{3}\right)\right]
                                   \nonumber \\[1mm]
W_{SA} &=& \hspace{3mm} Q^{2}\,|F_{4}|^{2}\nonumber\\[1mm]
W_{SB} &=&  \hspace{3mm}2\sqrt{Q^{2}}\left[
            x_{1}\,\mbox{Re}\left(F_{1}F^{\ast}_{4}\right)
           +x_{2}\,\mbox{Re}\left(F_{2}F^{\ast}_{4}\right)
             \right]\nonumber\\[1mm]
W_{SC} &=&- 2\sqrt{Q^{2}}\left[
            x_{1}\,\mbox{Im}\left(F_{1}F^{\ast}_{4}\right)
           +x_{2}\,\mbox{Im}\left(F_{2}F^{\ast}_{4}\right)
             \right]\nonumber \\[1mm]
W_{SD} &=& \hspace{3mm} 2\sqrt{Q^{2}}x_{3}\left[
            \,\mbox{Re}\left(F_{1}F^{\ast}_{4}\right)
           -\,\mbox{Re}\left(F_{2}F^{\ast}_{4}\right)
             \right]\nonumber \\[1mm]\nonumber
W_{SE} &=& -2\sqrt{Q^{2}}x_{3}\left[
            \,\mbox{Im}\left(F_{1}F^{\ast}_{4}\right)
           -\,\mbox{Im}\left(F_{2}F^{\ast}_{4}\right)
             \right]\\[1mm]\nonumber
W_{SF} &=&- 2\sqrt{Q^{2}}x_{4}\,\mbox{Im}\left(F_{3}F^{\ast}_{4}\right)
                             \\[1mm]\nonumber
W_{SG} &=& -2\sqrt{Q^{2}}x_{4}\,\mbox{Re}\left(F_{3}F^{\ast}_{4}\right)
                             \nonumber
\end{eqnarray}
The variables $x_i$ are defined by
$
%   \begin{eqnarray}
x_{1}= V_{1}^{x}=q_{1}^{x}-q_{3}^{x},\,
x_{2}= V_{2}^{x}=q_{2}^{x}-q_{3}^{x},\,
x_{3}= V_{1}^{y}=q_{1}^{y}=-q_{2}^{y},\,
x_{4}= V_{3}^{z}=\sqrt{Q^{2}}x_{3}q_{3}^{x},\,
$
%   \end{eqnarray}
where $q_i^{x}$ ($q_i^{y}$) denotes
the $x$ ($y$) component of the momentum of
meson $i$ in the hadronic rest frame.
They can easily be expressed in terms of $s_1$, $s_2$ and $s_3$
\cite{km1,km2}.

Note that the first 9 structure functions
originate from the hadronic system in a spin one state
($W_A, W_C, W_D, W_E$ from the axial vector current,
 $W_B$ from the vector current and $W_F, W_G, W_H, W_I$ from the
interference of the axial vector and vector current).
$W_{SA}$ originates only from a hadronic system in a spin zero
state
and the remaining six structure functions are interference terms
between the spin one and spin zero states.

An inspection of Eq.~(\ref{wi}) shows also that the structure functions
$W_E, W_F, W_H, W_{SC}, W_{SE}, W_{SF}$
require nontrivial phases of the amplitudes resulting from final state
interactions. Only  the $T$-odd correlations
$\bar{L}_X W_X ,\,\, X\in\{E,F,H,SC,SE,SF\}$
allow in principle for a measurement of
CP violating effects in the hadronic matrix elements (see next section).

The leptonic coefficients
$\bar{L}_X$ depend  on the two angles $\beta, \gamma$ and on
$\gamma_{VA}$:
  \begin{equation}
  \begin{array}{lcrllcrl}
\bar{L}_{A} &=& 1/2 & K_1 (1+\cthz) + K_2 \>;
&\hspace{5mm} \bar{L}_{SA}&=&     &K_2           \>;\\[1mm]
\bar{L}_{B} &=&     &K_1 \sthz +  K_2\>;
&\hspace{5mm}\bar{L}_{SB}&=&    &K_2 \sth \cchi  \>;\\[1mm]
\bar{L}_{C} &=& -1/2&K_1 \sthz \czchi\>;
&\hspace{5mm}\bar{L}_{SC}&=&    & 0              \>;\\[1mm]
\bar{L}_{D} &=&  1/2&K_1 \sthz \szchi\>;
&\hspace{5mm}\bar{L}_{SD}&=&  -   &K_2 \sth \schi\>;\\[1mm]
\bar{L}_{E} &=&     &\gva \cth        \>;
&\hspace{5mm}\bar{L}_{SE}&=&  -   &0             \>;\\[1mm]
\bar{L}_{F} &=&  1/2& K_1 \szth \cchi\>;
&\hspace{5mm}\bar{L}_{SF}&=&  -   &K_2 \cth      \>;\\[1mm]
\bar{L}_{G} &=&  -  &\gva \sth \schi  \>;
&\hspace{5mm}\bar{L}_{SG}&=&    &0               \>;\\[1mm]
\bar{L}_{H} &=& -1/2&K_1 \szth \schi \>;
&                                    \\[1mm]
\bar{L}_{I} &=&  -  &\gva \sth \cchi  \>;
&                                     \\
  \end{array}
\label{ldef2}
  \end{equation}
The coefficients $K_i$ are defined in Eq.~(\ref{kdef}).
Note that the coefficients $\bar{L}_{SC}, \bar{L}_{SE}, \bar{L}_{SG}$ vanish if
only the two
Euler angles $\beta$ and $\gamma$ are considered.
It has been shown in \cite{km1} that in the case of a polarized $\tau$
(as it is the situation at LEP) one can use the
direction of the $\tau$ spin-vector in the lab to define a further
Euler angle $\alpha$. If this additional angle is considered,
all 16 coefficients $\bar{L}_X$ in Eqs. (\ref{ldef2}) are nonvanishing
enabling the measurement of all 16 structure functions $W_X$.\\
The coefficients $\bar{L}_{SC}$ $\bar{L}_{SE}$ are of particular
importance for the  detection of possible  CP violation originating from
a charged Higgs exchange (see below).

Numerical results for the nonvanishing structure functions in the
$3\pi$ decay mode are discussed in \cite{km1,km2}.
Furthermore, it has been   shown in
\cite{km1} that the technique of the structure functions  allows
for a model independent test of possible spin zero components
(parametrized by $F_4$) in the hadronic current by analyzing the
structure functions $W_{SB}$ and $W_{SD}$.
Note that
the $\cos\beta$ distribution
allows already for a model independent
separation of the axial-vector  and the vector current  contribution
in the decay modes with different mesons,
{\it i.e.} the structure functions $W_A$ and
$W_B$ in Eq.~(\ref{wi})
can be disentangled due to the different $\beta$ dependence of
$\bar{L}_A$ and $\bar{L}_B$.
Numerical results of the structure functions for
several three meson decay modes with  different mesons based on the
model in \cite{Dec93} are discussed in \cite{dm2}. A more detailed analysis
(including the full $Q^2$ and $s_i$
dependence of the structure functions)
based on the parameterization in \cite{fm1}
is in preparation \cite{fm2}.

%%%%%%%%%%%%%%%%%%%%%%%%%%%%%%%%%%%%%%%%%%%%%%%%%%%%%%
\section{CP VIOLATION EFFECTS }
%%%%%%%%%%%%%%%%%%%%%%%%%%%%%%%%%%%%%%%%%%%%%%%%%%%%%%
Currently CP violation has been experimentally observed only in the
$K$ meson system. The effect can be explained by a nontrivial complex
phase in the Kobayashi-Maskawa flavour mixing matrix.
However, the fundamental origin of this CP violation
is still unknown.
CP-odd correlations of the $\tau^-$ and $\tau^+$ decay products,
which originate from an electric dipole moment in the $\tau$ pair
production, have been  discussed in \cite{nachtmann}.
In this paper, we investigate the effects of possible
non-Kobayashi-Maskawa-type of CP violation, {\it i.e.} CP violation
effects beyond the Standard Model. Such effects could originate for
example from multi Higgs boson models \cite{mhiggs},
scalar leptoquark model \cite{lepto} or left-right symmetric models
\cite{lhmodels}.

Any possible  observation of these  CP violation effects
needs not only a CP-violating complex phase (parametrized as $\eta$
and $\chi$ below)
in the hadronic matrix elements but also the interference with a CP
conserving phase resulting from final state interactions.
Therefore, only the correlations involving
structure functions proportional to the imaginary part of the
form factors $F_i$ allow in principle for an observation of CP
violation effects by taking the difference of
$d\Gamma[\tau^-]-d\Gamma[\tau^+]$
of the corresponding $T$-odd correlations (see below).

In the two meson decay modes, the only
structure function which is sensitive to CP-violation effects
is  $W_{SG}$ in Eq.~(\ref{wsg})
[proportional to $\mbox{Im}\left[FF_4^*\right]$].
Unfortunately, this structure function is not observable if
only distributions of the angle $\beta$ are considered,
{\it i.e.} the coefficient $\bar{L}_{SG}$ vanishes.
However, $W_{SG}$ could in principle be measured by taking into account
additional distributions with respect to the $\tau$ spin vector
(assuming   polarized incident  beams).

CP violation effects in the $\tau\rightarrow 2\pi \nu$ decay mode
from the scalar sector (e.g. the multi Higgs boson models)
have recently been discussed in terms of ``stage-two spin correlation
functions'' in \cite{nelson1} and in the case of polarized
electron-positron beams at $\tau$ charm factories in \cite{tsaicp}.
In \cite{nelson1}, the decay products of the second tau decay are used
to define a $T$-odd correlation
whereas  the $\tau$ polarization (assuming  a polarized incident
electron beam) is used in $\cite{tsaicp}$ to define
a $T$-odd triple correlation. In fact, the correlations in
\cite{tsaicp}
are equivalent to the product
$\bar{L}_{SG}W_{SG}$ as discussed before in the two meson case, if
the angle $\alpha$ is defined with respect to the $\tau$ spin as
described after Eq.~(\ref{kdef}).

In the three meson case,
the structure functions
$W_E, W_F, W_H$, $W_{SC}, W_{SE}, W_{SF}$
in Eq.~(\ref{wi})
require nontrivial phases of the amplitudes resulting from final state
interactions.
Only  the $T$-odd correlations
$\bar{L}_X W_X ,\,\, X\in\{E,F,H,SC,SE,SF\}$
allow therefore in principle for a measurement of
CP violating effects in the hadronic matrix elements.
As can be seen from Eq.~(\ref{ldef2})
the coefficients $\bar{L}_{SC}, \bar{L}_{SE}, \bar{L}_{SG}$ vanish if
only the two Euler angles $\beta$ and $\gamma$ are considered.
However, the structure functions $W_E, W_F, W_H, W_{SF}$
can be measured
through the $\beta$ and $\gamma$ dependence encoded in the
coefficients $\bar{L}_E, \bar{L}_F, \bar{L}_H, \bar{L}_{SF}$.

Let us therefore parametrize possible CP violation effects
in the hadronic decay amplitudes by replacing Eqs.~(\ref{f1234})
by
\begin{eqnarray}
J^{\mu}(q_{1},q_{2},q_{3})
&=&
   \left[
            \left(
             V_1^\mu\, F_1
            +V_2^\mu\, F_2  \right)\,\,(1+\chi_A)
            +V_4^\mu\, F_4  \,\,  \left(1+\chi_A+\eta \right)
            \right. \nonumber\\
&& \left.
   \hspace{-2mm}
  +\,i\,V_3^\mu\,F_3\,\,(1+\chi_V)\,\,   \right] \label{vinew}
\end{eqnarray}
where $V_i^\mu$ are given in Eq.~(\ref{videf}).

The  term proportional to $\eta$ parametrizes the effect
of a possible charged Higgs
boson \cite{mhiggs}, whereas the complex numbers $\chi_A$ and $\chi_V$
parametrize any new physics that would arise from vector or scalar
boson exchange motivated by left-right symmetric models
\cite{lhmodels}. The Standard Model prediction is obtained from
Eq.~(\ref{vinew}) by setting $\chi$ and $\eta$ to zero.
Let us now assume that the complex numbers  $\chi_A, \chi_V$
and $\eta$ transform like
\begin{equation}
\chi_A  \stackrel{\mbox{CP}}{\longrightarrow}\chi_A^\ast;\,\,\,\,\,\,
\chi_V  \stackrel{\mbox{CP}}{\longrightarrow}\chi_V^\ast;\,\,\,\,\,\,
\eta    \stackrel{\mbox{CP}}{\longrightarrow}\eta^\ast.
\label{cptrans}
\end{equation}

The hadronic structure functions $\tilde{W}_X$, which include the new physics
effects parametrized by the numbers $\eta$ and $\chi$ are easily
obtained from Eq.~(\ref{wi}) using the transformation
\begin{eqnarray}
F_1 & \rightarrow &\tilde{F}_1= F_1 (1+\chi_A)\>,\label{f1t}\\
F_2 & \rightarrow &\tilde{F}_2= F_2 (1+\chi_A)\>,\\
F_3 & \rightarrow &\tilde{F}_3= F_3 (1+\chi_V)\>,\\
F_4 & \rightarrow &\tilde{F}_4= F_4 (1+\chi_A+\eta)\>.\label{f4t}
\end{eqnarray}
The hadronic structure functions are affected by the
sign change in the weak phases
under CP transformation as described in  Eq.~(\ref{cptrans}).
Note that the strong (complex) phases due to final state
interactions [given by Breit-Wigner propagators for the
two body resonances] are not changed,
because the strong interaction is invariant
under charge conjugation.
Besides of the sign change in the weak phases, the
structure functions $\tilde{W}_F,
\tilde{W}_G, \tilde{W}_H,
\tilde{W}_I, \tilde{W}_{SF}, \tilde{W}_{SG}$, which
originate from the interference of the axial
vector and vector current, change sign.
Furthermore,
the amplitude for the CP conjugated process
$\tau^+$
can be obtained from the results for $\tau^-$ by
reversing all momenta and spins of the particles.
Thus, $\cos\beta \rightarrow - \cos\beta$ and
$\gamma_{VA}=-\gamma_{VA}$.
%and the angular coefficients $\bar{L}_{E,F,H,SC,SE,SF}$
%change sign under CP operation.
CP invariance therefore  relates the differential decay rates
for $\tau^+$ and $\tau^-$ as:
\begin{equation}
d\Gamma[\tau^-](\cos\beta,\gva,\tilde{W}_X)
\stackrel{\mbox{CP}}{=}
d\Gamma[\tau^+](-\cos\beta,-\gva,a_X\tilde{W}_X)
\end{equation}
with $a_X = -1$ for
$X\in\{\tilde{W}_F, \tilde{W}_G,
\tilde{W}_H, \tilde{W}_I, \tilde{W}_{SF}, \tilde{W}_{SG}\}$ and
$a_X=1$ else.

If CP is not violated, the difference
$d\Gamma[\tau^-]-d\Gamma[\tau^+]$ should vanish.
{}From the $T$-odd correlations
$\bar{L}_X \tilde{W}_X ,\,\, X\in\{E,F,H,SC,SE,SF\}$, one can construct
CP-violating quantities by taking the difference of
these correlations for $\tau^-$ and $\tau^+$.
\begin{eqnarray}
\Delta_X &=& \frac{1}{2}
\left(
\bar{L}_X( \cos\beta,\, \gva)\, \tilde{W}_X[\tau^-]
-
\bar{L}_X(-\cos\beta,\,-\gva)\, a_X \tilde{W}_X[\tau^+]
 \right)\nonumber \\
&=& \bar{L}_X(\cos\beta,\,\gva)\,\left( \tilde{W}_X[\tau^-] -
\tilde{W}_X[\tau^+]\right)
\,\,\,\equiv\,\,\, \bar{L}_X\,\Delta{\tilde{W}_X}\nonumber\>,\\
\end{eqnarray}
where
\begin{equation}
\Delta{\tilde{W}_X} = \tilde{W}_X[\tau^-] - \tilde{W}_X[\tau^+]
\end{equation}
The nonvanishing  CP-violating differences can be
calculated from Eqs.~(\ref{wi},\ref{f1t}-\ref{f4t}) and  expressed in terms
of the form factors $F_i$ and the complex numbers $\chi_A, \chi_V$ and
$\eta$ as follows:
\begin{eqnarray}
\Delta{\tilde{W}_F} &=&  \hspace{3mm}
          2x_{4}\left[x_{1}\,\mbox{Re}\left(F_{1}F^{\ast}_{3}\right)
                    + x_{2}\,\mbox{Re}\left(F_{2}F^{\ast}_{3}\right)\right]
         \, \mbox{Im}\left(\chi_A-\chi_V+\chi_A\chi_V^\ast\right)
         \label{dwf}
         \>,\\[2mm]
\Delta{\tilde{W}_H} &=& \hspace{3mm}
         2x_{3}x_{4}\left[\,\mbox{Re}\left(F_{1}F^{\ast}_{3}\right)
                     -\,\mbox{Re}\left(F_{2}F^{\ast}_{3}\right)\right]
         \, \mbox{Im}\left(\chi_A-\chi_V+\chi_A\chi_V^\ast\right)
           \label{dwh}
         \>,\\[2mm]
\Delta{\tilde{W}_{SF}} &=&
           - 2\sqrt{Q^{2}}x_{4}\,\mbox{Re}\left(F_{3}F^{\ast}_{4}\right)
           \mbox{Im}\left(\chi_V-\chi_A-\eta
            +\chi_V(\chi_A^\ast+\eta^\ast) \right)
            \>.    \label{wsf}
\end{eqnarray}
An observed nonzero values for these
differences would signal a true CP-violation.
Note that all CP-violating differences are proportional to the
imaginary part $\eta$ and $\chi$.
Note also that $\Delta \tilde{W}_E$ vanishs, because
the form factors $F_1$ and $F_2$ multiply the same
complex weak phase.
Eqs.~(\ref{dwf},\ref{dwh}) show  that CP violation
effects parametrized by $\chi_A$ and $\chi_V$ are in principle
observable in a Tau-Charm Factory for three meson decay modes
with a nonvanishing vector (proportional to $F_3$) {\it and}
and axial vector current (proportional to $F_1, F_2$).
CP violation effects from a charged Higgs could  be detected
through $\Delta{\tilde{W}_{SF}}$ only for decay modes with a
nonvanishing vector current. Therefore, CP-violation tests in the
three pion decay mode are not possible, if only the decay distribution
with respect to the angles $\beta$ and $\gamma$ are taken into account.

As mentioned before, it has been shown in \cite{km1}
that in the case of a polarized $\tau$
one can use the
direction of the $\tau$ spin-vector in the lab to define a further
Euler angle $\alpha$. This additional angular dependence
allows in principle for the measurement of the two additional
CP-violating differences
\begin{eqnarray}
\Delta{\tilde{W}_{SC}} &=& 2\sqrt{Q^{2}}\left[
            x_{1}\,\mbox{Re}\left(F_{1}F^{\ast}_{4}\right)
           +x_{2}\,\mbox{Re}\left(F_{2}F^{\ast}_{4}\right)
             \right] \mbox{Im}\left(-\eta+\chi_A \eta^\ast \right)\>,
         \nonumber
         \\[2mm]
\Delta{\tilde{W}_{SE}}&=&  2\sqrt{Q^{2}}x_{3}\left[
            \,\mbox{Re}\left(F_{1}F^{\ast}_{4}\right)
           -\,\mbox{Re}\left(F_{2}F^{\ast}_{4}\right)
             \right] \mbox{Im}\left(-\eta+\chi_A \eta^\ast \right)\>.
           \nonumber
\end{eqnarray}
and hence for CP violation tests originating from a  charged Higgs
in the three pion decay mode.

The authors in \cite{koerner} studied the effects of $T$-odd triple
correlations (as derived in \cite{km1}) in the decay modes
$\tau\rightarrow K\pi\pi\nu$ and
$\tau\rightarrow KK\pi\nu$ using the  model for the hadronic
form factors as suggested in \cite{Dec93}.
They found that CP violation effects
in some extensions of the Standard Model  could be as big as 0.1\%.
CP violating effects in the $\tau\rightarrow 3\pi\nu$ decay mode have
also been discussed in \cite{hagiwara}.

%%%%%%%%%%%%%%%%%%%%%%%%%%%%%%%%%%%%%%%%%%%%%%%%%%%%%%
\section*{ACKNOWLEDGEMENTS }
%%%%%%%%%%%%%%%%%%%%%%%%%%%%%%%%%%%%%%%%%%%%%%%%%%%%%%
%
We would like to thank J.H. Kuehn for collaboration on part of the work
presented here.
The work of E. M. was supported in part
by the U. S. Department of Energy under
Grant No. DE-FG02-95ER40896.  Further support was provided by the
University of Wisconsin Research Committee, with funds granted by the
Wisconsin Alumni Research Foundation.
The work of M.F. has been supported in part
by the National Science Foundation Grant PHY-9218167.

%%%%%%%%%%%%%%%%%%%%% REFERENCES %%%%%%%%%%%%%%%%%%%%%%%%%%%%%%%%%%%%%%%%%%%%%%
%

\end{document}